\documentclass[aip, jcp,
amsmath,
amssymb,
reprint,groupedaddress]{revtex4-1}

\usepackage{graphicx}
\usepackage{dcolumn}
\usepackage{bm}
\usepackage{xcolor}

\usepackage[utf8]{inputenc}
\usepackage[T1]{fontenc}
\usepackage{mathptmx}

\begin{document}

\preprint{}

\title[]{Role of pre-ordered liquid in the selection mechanism of crystal polymorphs during nucleation}

\author{Sarath Menon}
\email{sarath.menon@rub.de}
\author{Grisell D\'{i}az Leines}
\author{Ralf Drautz}
\author{Jutta Rogal}
\email{jutta.rogal@rub.de}
\affiliation{Interdisciplinary Centre for Advanced Materials Simulation, Ruhr-Universit{\"a}t Bochum, 44801 Bochum, Germany}
\date{\today}

\begin{abstract}
We investigate the atomistic mechanism of homogeneous nucleation during solidification in molybdenum employing transition path sampling.  
The mechanism is characterized by the formation of a pre-structured region of high bond-orientational order in the supercooled liquid followed by the nucleation of the crystalline bulk phase within the center of the growing solid cluster.
This precursor plays a crucial role in the process, as it provides a diffusive interface between the liquid and crystalline core, which lowers the interfacial free energy and facilitates the nucleation of the bulk phase. 
Furthermore, the structural features of the pre-ordered regions are distinct from the liquid and solid phases, and preselect the specific polymorph that nucleates.  The similarity in the nucleation mechanism of Mo with that of metals that exhibit different crystalline bulk phases indicates that the formation  of a precursor is a general feature observed in these materials.  The strong influence of the structural characteristics of the precursors on the final crystalline bulk phase demonstrates that for the investigated system polymorph selection takes place in the very early stages of nucleation.

\end{abstract}


\maketitle 

\section{Introduction} \label{sec:introd}

Polymorph selection during crystallization is one of the most fundamental processes that plays an important role in several applications, ranging from the development of pharmaceutical drugs to the design of novel metallurgical materials.  The final bulk polymorph of a material may already be largely determined in the early stages of crystallization, and thus, fundamental knowledge of the nucleation process on the atomic scale is essential.
Nevertheless, as many systems exhibit complex mechanisms of crystallization, characterized by multiple steps and the competition of various  crystalline structures within small clusters, fundamental understanding of crystal nucleation processes and polymorph selection mechanisms remains elusive.~\cite{Sosso2016}

Classical nucleation theory (CNT)~\cite{Becker1935,Binder1987} provides a successful phenomenological
description of homogeneous nucleation,
but because of the simplifying approximations and its application on the mesoscale, CNT 
cannot 
capture quantitative details of nucleation mechanisms. A main assumption of CNT, the well known capillarity approximation, is that small, spherical clusters
are in the same thermodynamic phase as the bulk
and have a sharp interface with the surrounding liquid. However,
corrections to the interfacial free energy term that take into account the shape of the crystalline clusters~\cite{Horsch2008, Prestipino2014} and the finite size of a diffusive interface~\cite{Granasy1996, Prestipino2012} are often needed to match experimental data for timescales and activation energies. Furthermore, in contrast to the assumptions within CNT, the formation of polymorphic clusters with intermediate phases that differ from the final bulk structure have been observed in a vast number of examples,~\cite{Sosso2016} even for simple model systems such as Lennard-Jones fluids~\cite{Swope1990, TenWolde1995, Frenkel1996} and hard spheres.~\cite{Auer2001}

Recently, several studies have reported non-classical nucleation mechanisms characterized by the initial formation of pre-ordered regions in the liquid that act as precursors of the crystallization and the selected polymorphic structures.~\cite{Schilling2010,Kawasaki2011, Lechner2011, Russo2012, Lederer2014,  Desgranges2007, DiazLeines2017,DiazLeines2018} The clusters of pre-structured liquid are regions of either increased bond-orientational order~\cite{Russo2012, Russo2012a, Kawasaki2011, Schilling2010} or density~\cite{Lutsko2006, Vekilov2004, TenWolde1997, TenWolde1999}  that promote the emergence of crystallites within the center of the clusters by reducing the interfacial free energy.~\cite{Kawasaki2011, Russo2018}
The observation of pre-ordered regions in the melt has raised great interest in understanding the connection between structural and dynamical heterogeneity of the liquids and crystallization mechanisms.~\cite{Gasser2003, Jakse2003, Jakse2004b, Russo2018, Puosi2019}
Russo \emph{et al.} recently showed for model liquids
that the structural differences between liquids and crystals indeed control their glass-forming or crystal-forming ability, by suppressing or promoting the formation of crystalline precursors  via a thermodynamic interface penalty.~\cite{Russo2018}
Furthermore, a structural analysis of pre-ordered liquid regions in hard spheres~\cite{Kawasaki2011} showed that the structural features  of these regions resemble the coordination polyhedra of the crystalline structures formed in the growing clusters and therefore pre-determine the polymorphs selected during crystallization.
 Previously, we have shown that during solidification in the face-centered cubic (fcc) metal nickel, the pre-ordered liquid region
plays an essential role in the structural description of the growing nucleus and its interfacial free energy, and thus, represents an order parameter that significantly enhances the reaction coordinate.~\cite{DiazLeines2018}
 Moreover, in agreement with previous findings for hard sphere models, these regions of higher bond-orientational order than the liquid predetermine the coordination of the fcc-hexagonal closed packed (hcp) polymorphs selected in Ni, acting as precursors of the crystallization. But questions remain whether this crystallization mechanism is generally to be observed in other metallic systems that exhibit different thermodynamically stable phases, and how the different structural nature of the pre-ordered liquid determines the polymorph selected in the growing solid nucleus. 
 
 In this work, we take a step further to address these questions and investigate the nucleation process during solidification in a body-centered cubic (bcc) metal, molybdenum, using transition path sampling (TPS) simulations.~\cite{Dellago2002,VanErp2005}
 Molybdenum is widely used as a component in steel alloys as it improves corrosion resistance and weldability.~\cite{Garner1977} The nucleation process in Mo is largely unknown, partially due to the high melting point (2896~K) that hampers experimental studies.
In bcc metals, only a few theoretical studies exist, which focus on molecular dynamics (MD) simulations of rapid solidification during quenching in iron~\cite{Pan2015, Wang2018, Li2014} and zirconium,~\cite{Wu2011} where it was observed that icosahedral short-range order in the liquid gradually transforms into bcc-like short-range order during nucleation without the formation of other competing phases.
Likewise, Wang \emph{et al.}~\cite{Wang2018} showed that regions of  bcc-like short-range order yield the formation of bulk bcc during crystallization in Fe. However, due to the rare event nature of nucleation events, straightforward MD simulations are limited to extremely high cooling rates that can result in trajectories that strongly differ from the actual mechanism of crystallization at moderate undercoolings.
Here, we use the statistical path ensembles obtained from TPS to analyze the nucleation mechanism and kinetics in Mo at moderate undercoolings. We show that, similar to our findings in Ni,~\cite{DiazLeines2017, DiazLeines2018} the initial formation of  pre-ordered regions with increased bond-orientational order within the liquid promotes the emergence of 
the crystalline phase
within the core of these clusters, acting as a precursor of crystallization.    
At different undercoolings, the structural characteristics of the pre-ordered regions appear similar, whereas the frequency of formation and extend of pre-ordered regions is more pronounced at larger undercoolings. 
Our results suggest that the structural heterogeneity of the undercooled liquid, characterized by differences between regions with low and high bond-orientational order, is directly linked to crystal nucleation in Mo. The strong spatial correlation between regions of high bond-orientational order in the liquid and critical fluctuations also implies that the structure of the undercooled liquid is indeed decisive for the nucleation mechanism. 
We find that the overall nucleation mechanism is not sensitive to the employed interatomic potentials, provided the structural features of both, the solid and the liquid phase, are correctly captured,  further corroborating the relation between the structural heterogeneity of the liquid and crystallization. 
With an extensive analysis of the Voronoi polyhedra found in the pre-structured liquid regions in Mo, we show that the structural features of these regions  are inherently different from those found in crystalline precursors of fcc metals. 
The correspondence between pre-structured regions within the liquid and locally favored bcc-like ordering strongly indicates that the precursors predetermine the final bulk phase.
In these materials, the selection of the bulk polymorph thus takes place in the early stages of the crystallization process.

\section{Computational Approach}
\label{sec:methods}

\subsection{Simulation setup}
\label{subsec:tps}

To investigate the initial stages of nucleation and growth during solidification in molybdenum we employ transition interface sampling~(TIS).~\cite{VanErp2003}  TIS is a variant of transition path sampling~(TPS)~\cite{Dellago2002} in which an ensemble of trajectories is created that connects two (meta-)stable states in phase space.  In the current study, the two states of interest correspond to the liquid and solid phases in Mo.  The ensemble of pathways is sampled with a Monte Carlo (MC) procedure in trajectory space, and the analysis of the path ensemble provides access to both kinetic and thermodynamic properties of the transition.~\cite{VanErp2005}

As an order parameter (or collective variable (CV)) to distinguish between the solid and liquid state we use the size of the largest cluster consisting of solid particles, $n_s$, introduced in section~\ref{subsec:simulationdetails}.  The same order parameter is also used to define the positions of the interfaces in the TIS simulations.
The TIS path ensembles are subsequently reweighted to obtain an estimate of the complete ensemble~\cite{Rogal2010,Bolhuis2011} that contains each trajectory with its correct weight to represent unbiased simulations.  It is then possible to project the reweighted path ensemble into different low-dimensional CV spaces and compute quantities such as the free energy or averaged committor along arbitrary CVs.~\cite{Bolhuis2011}
The free energy, $F$, in a $m$-dimensional CV space $\mathbf{q} = \{q_1(\mathbf{x}),\dots,q_m(\mathbf{x})\}$, where $\mathbf{x}$ is a point in phase space, is, for example, given by
\begin{equation}
 \label{eq:freeEprojected}
F(\mathbf{q}) = -k_\text{B} T \ln \rho(\mathbf{q}) + \text{const.}
\end{equation}
with the Boltzmann constant $k_\text{B}$, the temperature $T$, and the probability density $\rho(\mathbf{q})$, which is directly obtained from the reweighted path ensemble (RPE)~\cite{Rogal2010}
\begin{equation}
\label{eq:rhoRPE}
\rho(\mathbf{q})= C\int\mathcal{D} \mathbf{x}^L \mathcal{P}[\mathbf{x}^L] \sum_{k=0}^L \delta(\mathbf{q}(\mathbf{x}_k)-\mathbf{q}) \quad .
\end{equation}
Here, $\mathcal{P}[\mathbf{x}^L]$ is the RPE, $\int\mathcal{D} \mathbf{x}^L$ denotes the integral over all paths of length $L$, $\mathbf{x}^L=\{\mathbf{x}_0,\dots,\mathbf{x}_L\}$, 
where $\mathbf{x}_i$ are slices along the path,
$\delta(\mathbf{z}) = \prod_{i=1}^m \delta(z_i)$ is the Dirac delta function, and $C$ is a normalization constant.

All molecular dynamics (MD) trajectories sampled in the TIS simulations were generated using the {\scshape lammps}~\cite{Plimpton1995} code.  The simulation box contained $N=4394$ atoms which was found to be sufficient to minimize finite size effects.  To describe the interactions between Mo atoms an embedded atom method~(EAM) potential~\cite{Zhou2004} was used.  The melting temperature of this potential was determined with the z-method~\cite{Belonoshko2006} to be $T_m = 3472$~K.  All TIS simulations were performed in the isothermal-isobaric~(NPT) ensemble using a Nos\'{e}-Hoover thermostat and barostat as implemented in  {\scshape lammps} with an integration times step of $\Delta t = 2$~fs.  The temperature was set to 2592~K, 2708~K, and 2782~K, corresponding to 25\%, 23\%, and 20\% undercooling, respectively, at a pressure of $P=0$.
The TIS simulations were performed using a python wrapper~\cite{tpswrapper} with {\scshape lammps} as the MD driver.
The MC moves to create new trajectories during the simulations include shooting (65\%) as well as exchange between interface ensembles (35\%) to improve ergodicity of the sampling.~\cite{VanErp2007}
Trajectories were recorded every five MC steps to ensure sufficient decorrelation.  Along the MD trajectories, configurations were stored every 100~fs at 25\% undercooling and every 200~fs at 23\% and 20\% undercooling.
For each interface a total number of 900 trajectories was included in the path ensemble.

\subsection{Structure analysis}  \label{subsec:simulationdetails}

\begin{figure}
\includegraphics[width=0.4\textwidth]{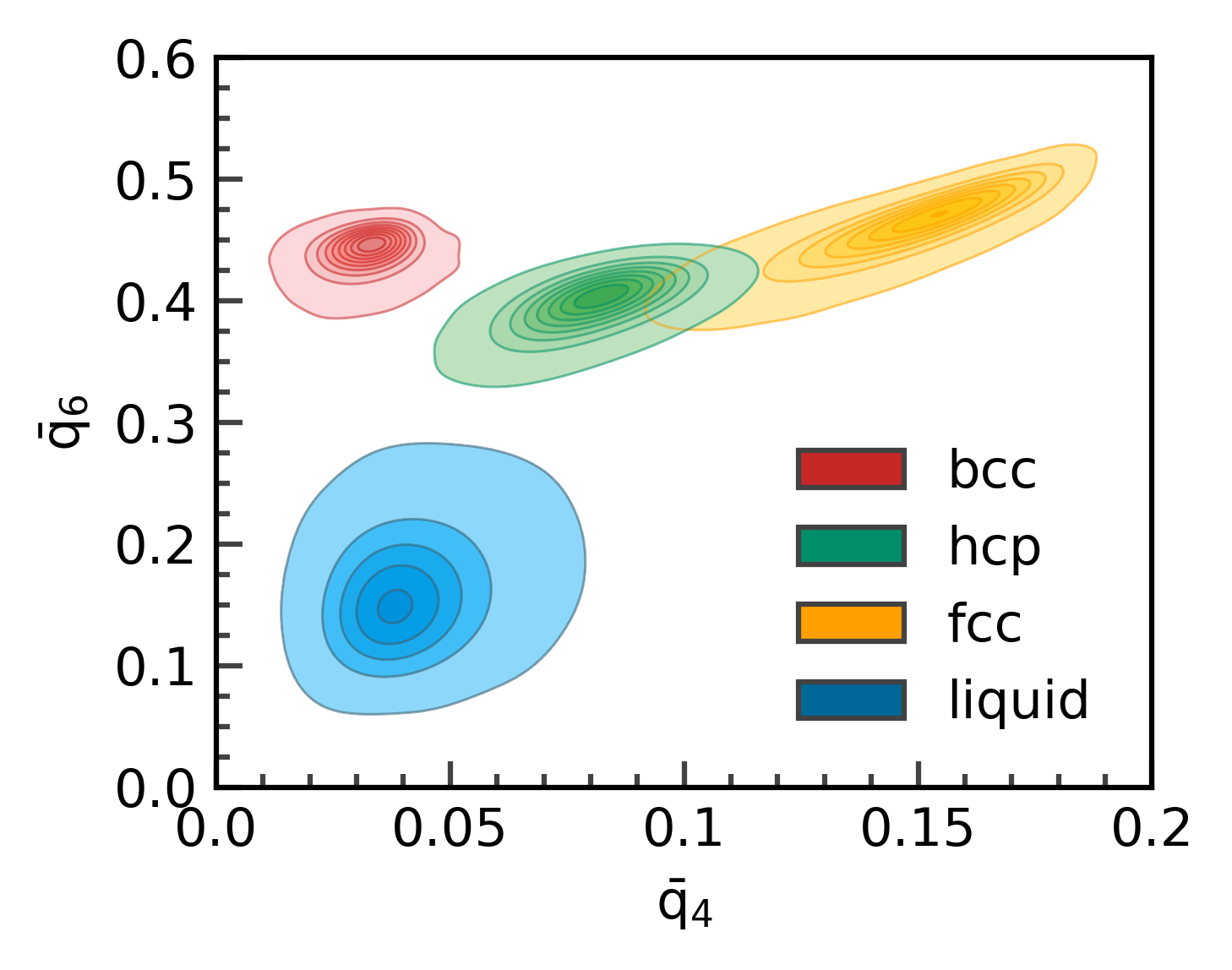}
\caption{\label{fig:q4q6map}
Probability distribution of the averaged local bond order parameters $\bar{q}_4$ and $\bar{q}_6$ for bcc~(red),  hcp~(green), fcc~(yellow), and liquid~(blue) in Mo at 20\% undercooling. A darker shade of the colors indicates a higher probability density.}
\end{figure}

A key component in the analysis of the simulation results is the identification of the local structure around each atom.  As a first step, we need to distinguish if an atom is in a liquid or solid environment.  Here, we use a criterion that determines the structural correlation of each atom with its neighbors based on the Steinhardt bond order parameters.~\cite{Steinhardt1983,Auer2005}  A solid bond between two atoms $i$ and $j$ exists if the correlation  $s_{ij} = \sum_{m=-6}^{m=6}q_{6m}(i)q_{6m}^*(j) > 0.5$, where $q_{lm}$ are the complex vectors of the spherical harmonics.
As a second criterion, we determine the average correlation over the nearest neighbors~\cite{Bokeloh2011,DiazLeines2017} $\langle s_{i} \rangle = 1/N_\text{nn} \sum_{j=1}^{N_\text{nn}} s_{ij}$ which refines the classification of solid atoms at the solid-liquid interface.
An atom is considered as solid if it has more than seven solid bonds and if $\langle s_{i} \rangle > 0.6$.
Using a clustering algorithm we can then identify clusters of solid atoms where $n_s$ represents the number of atoms in the largest solid cluster.

Specific crystal structures are assigned to solid atoms using the averaged local bond order parameters,~\cite{Lechner2008} $\bar{q}_4$ and $\bar{q}_6$.  The corresponding reference map for Mo at 20\% undercooling is shown in Fig.~\ref{fig:q4q6map}.
To compute the reference maps, MD simulations were performed in the NPT ensemble for bcc, fcc, and hcp bulk structures as well as for the liquid phase for 3~ns at each undercooling. The simulation box contained 6912 atoms for fcc and hcp, and 3456 atoms for bcc and liquid.  For each reference structure 300 configurations were randomly chosen from the MD trajectories and the $\bar{q}_4$ and $\bar{q}_6$ values were calculated using the {\scshape pyscal} library.~\cite{Menon2019}
In the analysis of the TIS simulations, an atom is assigned to a particular reference structure if the probability from the $\bar{q}_4 - \bar{q}_6$ distribution is larger than 0.01.  A solid particle that does not fall into the region of any of the crystalline structures on the  $\bar{q}_4 - \bar{q}_6$ map is defined as \emph{pre-ordered}.


\section{Nucleation and growth in undercooled Mo}
\label{sec:Mechanism of nucleation}

\subsection{Mechanism of nucleation}

\begin{figure*}
\includegraphics[width=1\textwidth]{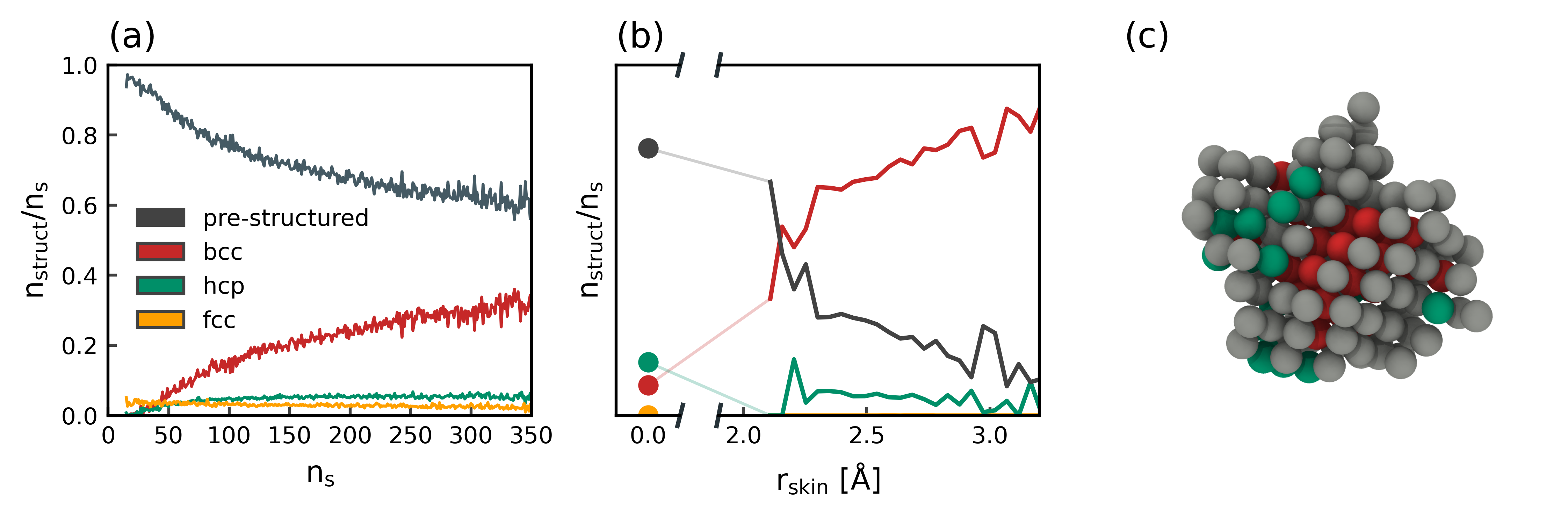}
\caption{\label{fig:struc_composition}
(a) Average structural composition of the growing nucleus at 20\% undercooling
(b) Fraction of crystalline structures and pre-ordered liquid as a function of the minimum distance to the surface of the solid cluster, $r_\text{skin}$, for critical nuclei at 20\% undercooling. There are no data at distances up to $\approx$ 2 \AA~from the surface atoms as this corresponds to the thickness of the surface layer. Transparent lines are a guide to the eye.
(c) Snapshot of a typical configuration of the critical nuclei. 
}
\end{figure*}

A basic step in investigating the nucleation mechanism is the analysis of the structural composition of the growing solid cluster, which can be directly extracted from the TIS path ensemble.
The structural composition of the largest solid cluster, $n_s$, as an average over 600 trajectories connecting the liquid and solid state at 20\% undercooling is shown in Fig.~\ref{fig:struc_composition}~(a). 
Similarly, the structural evolution was analyzed for 23\% and 25\% undercooling.
 For all undercoolings, the largest solid cluster is composed of about 85\% pre-structured  particles up to approximately half of the critical cluster size. Subsequently,  bcc emerges 
 while the pre-ordered particles continue to constitute a significant fraction of the cluster. Both fcc and hcp do not play any noticeable role during the entire nucleation process in Mo. We further determine the spatial distribution of different crystal structures within the critical solid clusters, that is at the transition state, for different undercoolings. 
 The distribution 
 at 20\% undercooling ($n^*_s=172$) averaged over 300 configurations is shown in Fig.~\ref{fig:struc_composition}~(b).  
 Here, $r_\text{skin}$ indicates the minimum distance to the surface of the solid cluster. The surface is defined by the atoms in the solid cluster that have at least one liquid neighbor. At the surface, where $r_\text{skin} = 0$~\AA, the solid cluster is predominantly composed of pre-ordered liquid particles whereas the core region consists of mainly bcc.  A representative snapshot of a critical nucleus is shown in Fig.~\ref{fig:struc_composition}~(c). 
The nucleation mechanism in Mo thus comprises the formation of pre-structured regions followed by the nucleation of the thermodynamically stable, bcc bulk phase within the centers of these regions. For all three undercoolings studied in this work, the overall mechanism remains the same, except for small differences in the fraction of bcc  that slightly decreases at higher undercoolings. 
This is predominantly due to kinetic effects, where shorter timescales at larger undercoolings hinder the rearrangement into the thermodynamically stable phase.
Similar nucleation mechanisms have also been observed in the fcc metals Al~\cite{Desgranges2007} and Ni~\cite{DiazLeines2018}, exhibiting the formation of pre-structured liquid regions followed by the emergence of a crystalline core.  
The evolution of the structural composition in these fcc metals is, however, different, as they exhibit random hcp stacking together with fcc, whereas in bcc Mo other crystalline structures do not play a role.

\begin{figure}
\includegraphics[width=0.36\textwidth]{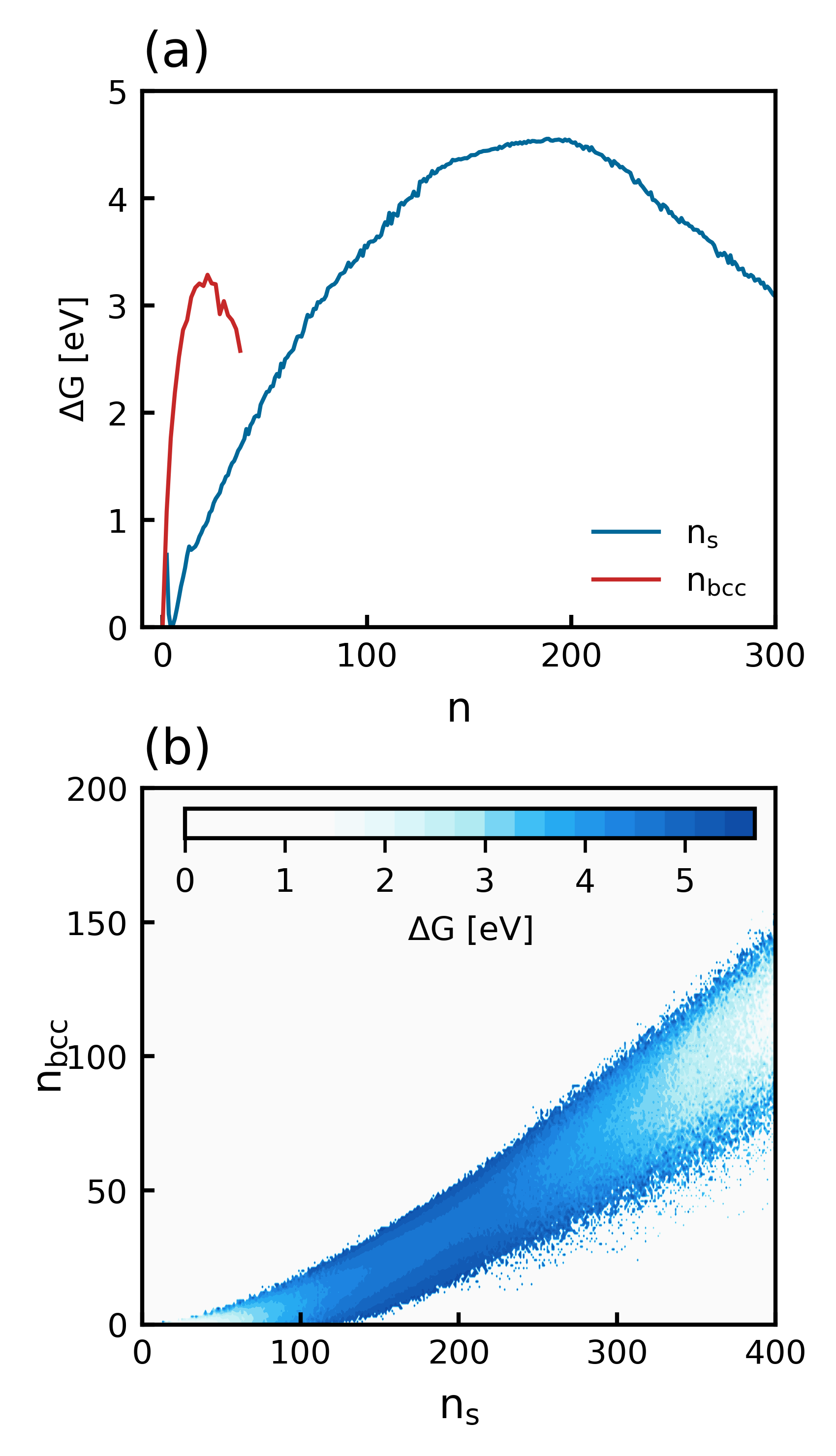}
\caption{ \label{fig:fes1d-2d}
(a) Free energy profiles projected from the reweighted path ensemble onto the size of the largest cluster, $n_s$~(blue), and on the number of bcc atoms in the largest cluster, $n_\text{bcc}$~(red), for 20\% undercooling.
(b) 2-dimensional free energy projection onto $n_s$ and $n_\text{bcc}$.
}
\end{figure}

In addition to the analysis of the structural evolution during nucleation, we assess  the free energy profile projected from the RPE according to Eq.~\eqref{eq:freeEprojected}.
In Fig. \ref{fig:fes1d-2d} (a) the blue line represents the free energy as a function of the largest solid cluster, $n_s$, at 20\% undercooling.  
The nucleation barrier is extracted from the maximum along the free energy profile, yielding 4.4~eV at 20\% undercooling.  The barrier decreases to 3.7~eV and 2.1~eV at 23\% and 25\% undercooling, respectively. 
Furthermore, the nucleation rates can directly be computed from the TIS ensemble. The associated timescales for nucleation vary from microseconds to nanoseconds with increasing undercooling (details concerning rates and barriers are given in the Supplementary Material~\cite{supplemental}).
The decrease in free energy barriers and increase in nucleation rates at higher undercoolings are in qualitative agreement with CNT. 

The initial formation of a pre-structured region in the liquid has a significant impact on the nucleation barrier. 
If we project the free energy along the number of bcc particles in the largest cluster, $n_\text{bcc}$ (which would be the parameter of choice within the capillarity assumption of CNT), the barrier decreases by more than 1~eV, see red line in Fig.~\ref{fig:fes1d-2d} (a).  In addition, the critical nucleus size decreases dramatically.
This is due to the fact that a projection onto $n_\text{bcc}$ cannot capture the initial pre-ordering in the liquid, which subsequently serves as a precursor for the nucleation of bcc.
The 2-dimensional free energy projection onto $n_s$ and $n_\text{bcc}$ shown in Fig.~~\ref{fig:fes1d-2d}~(b) further supports our interpretation.
Up to half of the critical cluster size, during the initial stage of the nucleation, $n_\text{bcc}$ is close to zero indicating the absence of bcc particles in the cluster. Consequently, the initial step is dominated by clustering of pre-structured particles, accompanied  by an increase in the free energy of about 1~eV.
When projecting only onto the number of bcc atoms, $n_\text{bcc}$, this initial increase due to pre-ordering is not captured, leading to an overall smaller nucleation barrier. 
The same effect was observed for all undercoolings studied in this work.

The initial stage of pre-ordering has an important role in the nucleation mechanism as it acts as a diffusive interface between the liquid and crystal core, facilitating crystal nucleation, and thus acting as a precursor. 
This is similar to the findings for fcc Ni~\cite{DiazLeines2018},  where including the pre-ordered region in the CV was necessary to accurately determine the free energy barrier. Comparing the nucleation mechanisms in bcc Mo and fcc Ni, we observe strong similarities: the initial step is characterized by the formation of a pre-ordered liquid state that acts as a precursor and is already associated with an increase in the free energy.
A crystalline core emerges within that region, surrounded by an interface layer of pre-structured liquid that persists throughout the nucleation process. 
However, the structure of the pre-ordered liquid is inherently different in the two systems, as discussed in the following section.

\subsection{Structural features of the pre-ordered liquid}

The pre-ordered liquid state observed during the nucleation is not just a short-lived fluctuation in the liquid, but a mesocrystal phase with a structure that is different from both the liquid and the bulk phase.  The structural characteristics of the pre-ordered liquid can already point towards the crystalline bulk phase that will nucleate from it.  Three aspects are of particular interest: the density, the crystallinity, and the local coordination polyhedra.

\subsubsection{Density}

\begin{figure}
    \includegraphics[width=7.0cm]{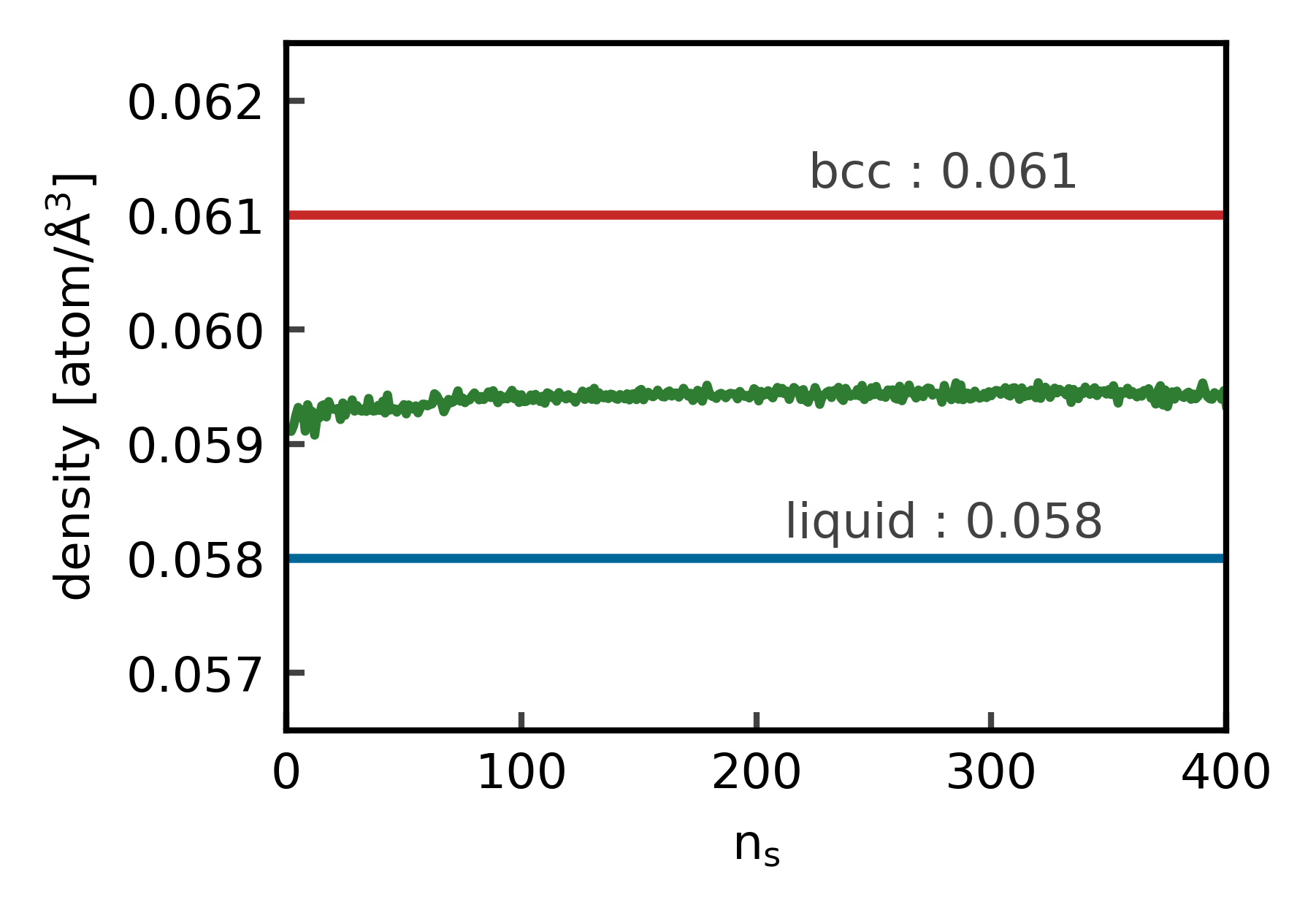}
    \caption{\label{fig:rho-eval} Average atomic density of the pre-structured particles
    (green) as a function of $n_s$, together with the inverse volumes of bcc (red) and liquid (blue).
}
\end{figure}

The atomic density is obtained from the inverse of the Voronoi volume computed using Voro++.~\cite{Rycroft2009}
At all undercoolings the difference in average density between the solid and the liquid is  small (0.003~atom/\AA$^3$ at 20-25\% undercoolings), 
as expected for a metallic bcc system.
In Fig.~\ref{fig:rho-eval} the average atomic density of the pre-structured particles in the largest solid cluster is shown as a function of $n_s$ extracted from 600 liquid to solid trajectories at 20\% undercooling. For each atomic configuration with a particular value of $n_s$, pre-structured atoms that belong to the solid cluster with at least two other pre-structured particles as neighbors are included in the calculation of the density.
Results for other undercoolings are comparable.
The density of the pre-ordered liquid particles is slightly larger than the one of the liquid indicating that the pre-ordered regions are more compact and distinct from the liquid phase. Since no change in the density of the pre-structured region is observed as the nucleation progresses, the density itself does not appear to drive the transformation process.
The formation of a precursor can be triggered by fluctuations either in the density~\cite{Lutsko2006, Vekilov2004, TenWolde1997, TenWolde1999} or the bond orientational order.~\cite{Russo2012, Russo2012a, Kawasaki2011, Schilling2010} In the case of Mo the small change in density can be considered rather as a result than a cause of the pre-ordering.

\subsubsection{Crystallinity}

\begin{figure*}
	\includegraphics[width=1.0\textwidth]{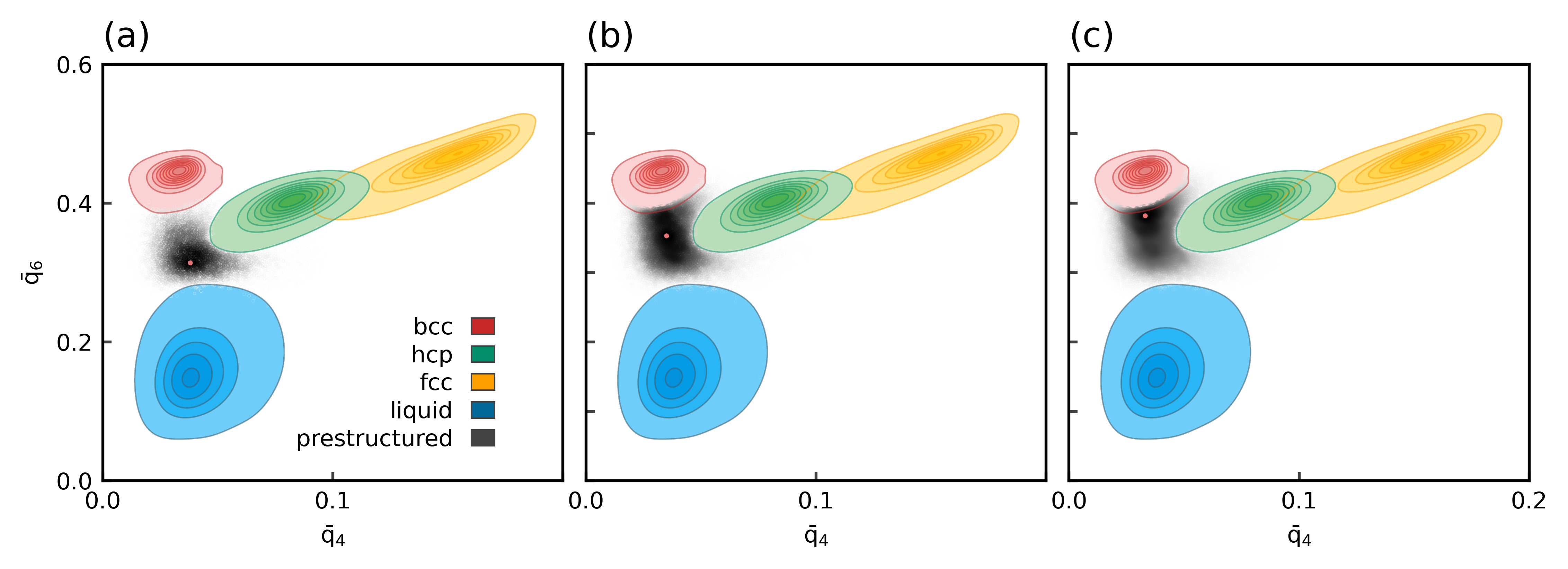}
	\caption{\label{fig:pre_structured}   Distributions of pre-structured particles~(black) on the $\bar{q}_4-\bar{q}_6$-map for (a) pre-critical, $n_s = 30$, (b) critical, $n_s^{*} = 172$ and (c) post-critical, $n_s = 370$~(right) clusters. Distributions of bulk bcc~(red), hcp~(green), fcc~(yellow), and liquid~(blue) are also shown. Regions of higher probability are indicated with a darker shade of the color.
The peak of the distribution of pre-structured particle (red dot) shifts from (0.038, 0.314) in the pre-critical state to (0.035, 0.353) in the critical cluster and to (0.033, 0.382) for post-critical clusters.
}
\end{figure*}

The degree of orientational order in the pre-structured liquid is evaluated by computing the averaged bond order parameters $\bar{q}_4$ and $\bar{q}_6$.  In Fig.~\ref{fig:pre_structured} the distribution of pre-ordered particles on the $\bar{q}_4-\bar{q}_6$-map is presented averaged over 300 configurations each with pre-critical, critical, and post-critical cluster sizes at 20\% undercooling.  
The analysis at 23\% and 25\% undercooling yields similar results.
The distributions occupy the region between the liquid and the crystalline (bcc, hcp, fcc) phases, again clearly distinct from either of them.  Compared to the liquid, the bond orientational order is increased, but not quite crystalline yet, similar to results for the pre-structured cloud observed during nucleation in Ni.~\cite{DiazLeines2017}  However, in contrast to the nucleation in an fcc material the distributions of pre-structured atoms are shifted towards the bcc region.  In particular, as the cluster size increases the peak of the distribution (indicated by the red dot in Fig.~\ref{fig:pre_structured}) moves closer to bcc, that is the orientational order in the pre-structured state continues to strengthen.
While the density remains constant the orientational order increases during nucleation and growth, thus promoting the formation of the bcc phase.
The location of the distribution of the pre-ordered state on the $\bar{q}_4-\bar{q}_6$-map indicates that the selection of the final bulk polymorph is already triggered in the very early stages of nucleation by the structural features of the precursor.

\subsubsection{Voronoi polyhedra}

In addition to the bond orientational order we analyze the local coordination around each atom by determining the corresponding Voronoi polyhedra (VP).
The VP are characterized by a quartet of integers $\langle n_3, n_4, n_5, n_6 \rangle$, where $n_i$ denotes the number of faces with $i$~vertices.~\cite{Finney1970,Tanemura1977}
In general, a single Voronoi polyhedron around an atom is not unique with respect to a specific crystal structure or phase, but each structure has a specific distribution of VP that acts as fingerprint.
The most common VP that are found during nucleation and growth of bcc can  be categorized broadly into four groups: $\langle~0~0~12~0 \rangle$ for a perfect icosahedron, $\langle~0~1~10~x\rangle$ for distorted icosahedra, $\langle~0~6~0~8~\rangle$ for bcc, and a wide range of polyhedra of coordination numbers of 13 and 14 for distorted bcc.

\begin{figure*}
	\includegraphics[width=1\textwidth]{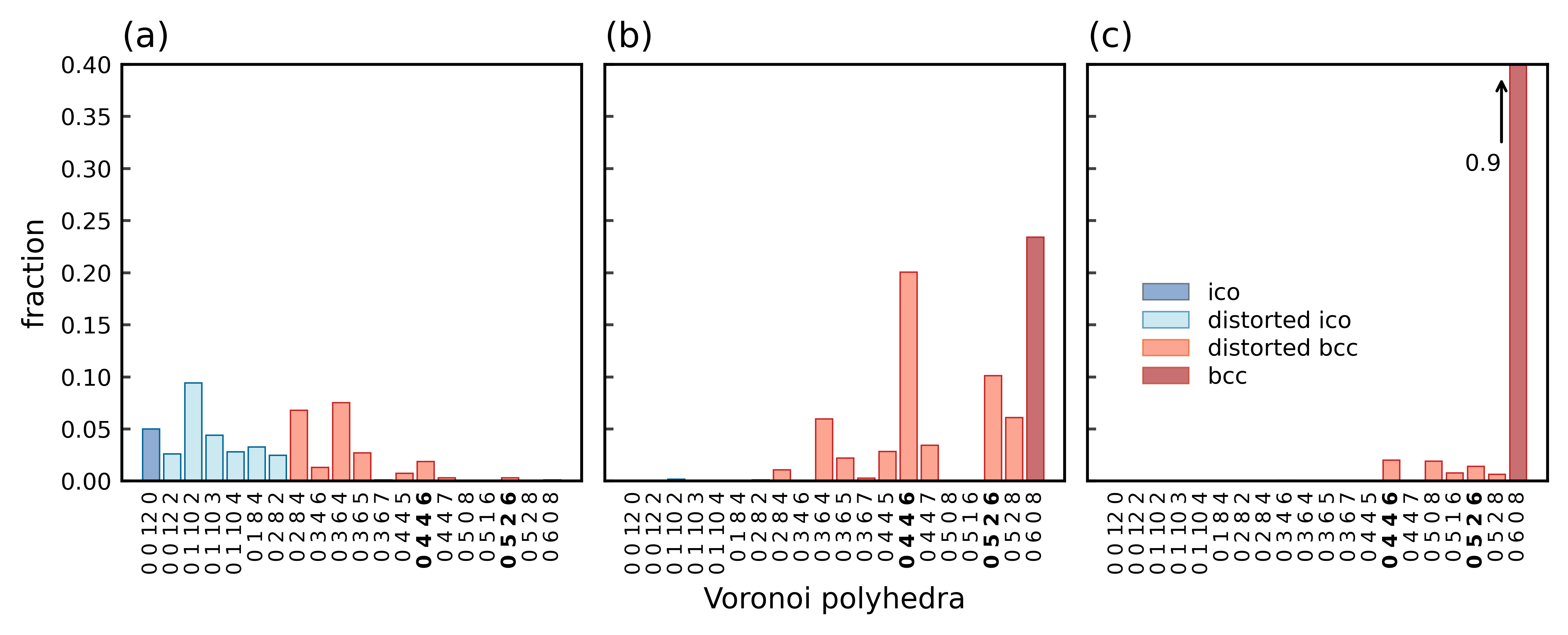}
	\caption{\label{fig:bulk_voronoi}  Distribution of Voronoi polyhedra for (a) liquid, (b) pre-structured liquid and (c) solid bcc at 20\% undercooling. The distributions are shown for VP that include the top 10 polyhedra most commonly found in the liquid, pre-ordered region and solid.  Icosahedral polyhedra are shown in dark blue, distorted icosahedral polyhedra are in light blue, distorted bcc and bcc are in light red and red respectively. The most common distorted bcc polyhedra, $<0~4~4~6>$ and $<0~5~2~6>$ are highlighted in bold. $<0~6~0~8>$ polyhedra in bcc peaks at a value of 0.9. }
\end{figure*}

In Fig.~\ref{fig:bulk_voronoi} we compare the distribution of VP at 20\% undercooling extracted from 100 configurations in the liquid, the pre-structured cloud at the critical nucleus size, and in bulk bcc.
In the case of undercooled liquid, we focus on the \emph{inherent} structure and relax the corresponding configurations until all forces are below $10^{-4}$~eV/\AA{} before determining the VP. 
For the pre-structured region and the solid we calculate the VP directly from the configurations.
Additionally, we do not consider faces that contribute less than 1\% to the total area in the classification of the VP.~\cite{Stukowski2012}

In the supercooled liquid a very broad distribution of VP is found.  The ten most common ones include the perfect icosahedron (dark blue in Fig.~\ref{fig:bulk_voronoi}), a number of distorted icosahedra (light blue), and a few distorted bcc polyhedra (light red), similar to the findings in Fe~\cite{Wang2018} and Zr.~\cite{Wu2011}
In bulk bcc mostly $\langle~0~6~0~8~\rangle$ VP (dark red) with minor amounts of distorted bcc  polyhedra are observed.  The distribution of VP in the pre-structured liquid features mostly distorted bcc polyhedra, whereas icosahedra and distorted icosahedra disappear. Thus, coming from the liquid a clear shift of the VP distribution takes place for the pre-ordered atoms.  This corroborates the fact that the local structural environments in the pre-ordered cloud are very different from liquid.
Furthermore, the VP in the pre-structured liquid show a clear tendency towards bcc ordering.
Polyhedra of the type $\langle~0~6~0~8~\rangle$, although distorted as compared to the ones in perfect bulk bcc, appear frequently in the pre-structured liquid.
The other two most common polyhedra of type $\langle~0~4~4~6~\rangle$ and $\langle~0~5~2~6~\rangle$ (bold labels in Fig.~\ref{fig:bulk_voronoi}) have previously been suggested as primary precursors for the formation of bcc Fe during quenching.~\cite{Pan2015}  Likewise, these VP are dominant in the distribution for the pre-structured atoms, but do not play a major role in either the liquid or bulk bcc.

The analysis of the VP suggests that the pre-ordered liquid provides a precursor for the emergence of a specific bulk polymorph. Icosahedral motifs that generally hamper the formation of crystalline phases are reduced and local atomic environments with higher bond orientational order already reflect characteristic features of the final bulk phase.  This indicates that in metallic systems polymorph selection takes place in the very early stages of nucleation during the formation of the precursor zones.

In the range of undercoolings we have considered, from 20-25\%, the structural characteristics of the pre-ordered liquid are very similar.  The distribution of VP for the pre-structured atoms shows a small decrease in the bcc polyhedron $\langle~0~6~0~8~\rangle$ from 34 to 27\% and respective increase in the two most important, distorted bcc polyhedra $\langle~0~4~4~6~\rangle$ and $\langle~0~5~2~6~\rangle$ with increasing undercooling. The slightly reduced structural ordering at larger undercoolings stems from the faster nucleation kinetics, as the atoms have less time to arrange into the thermodynamically preferred state. 

\subsection{Influence of interatomic potentials on nucleation mechanism}

\begin{figure*}
    \includegraphics[width=12.0cm]{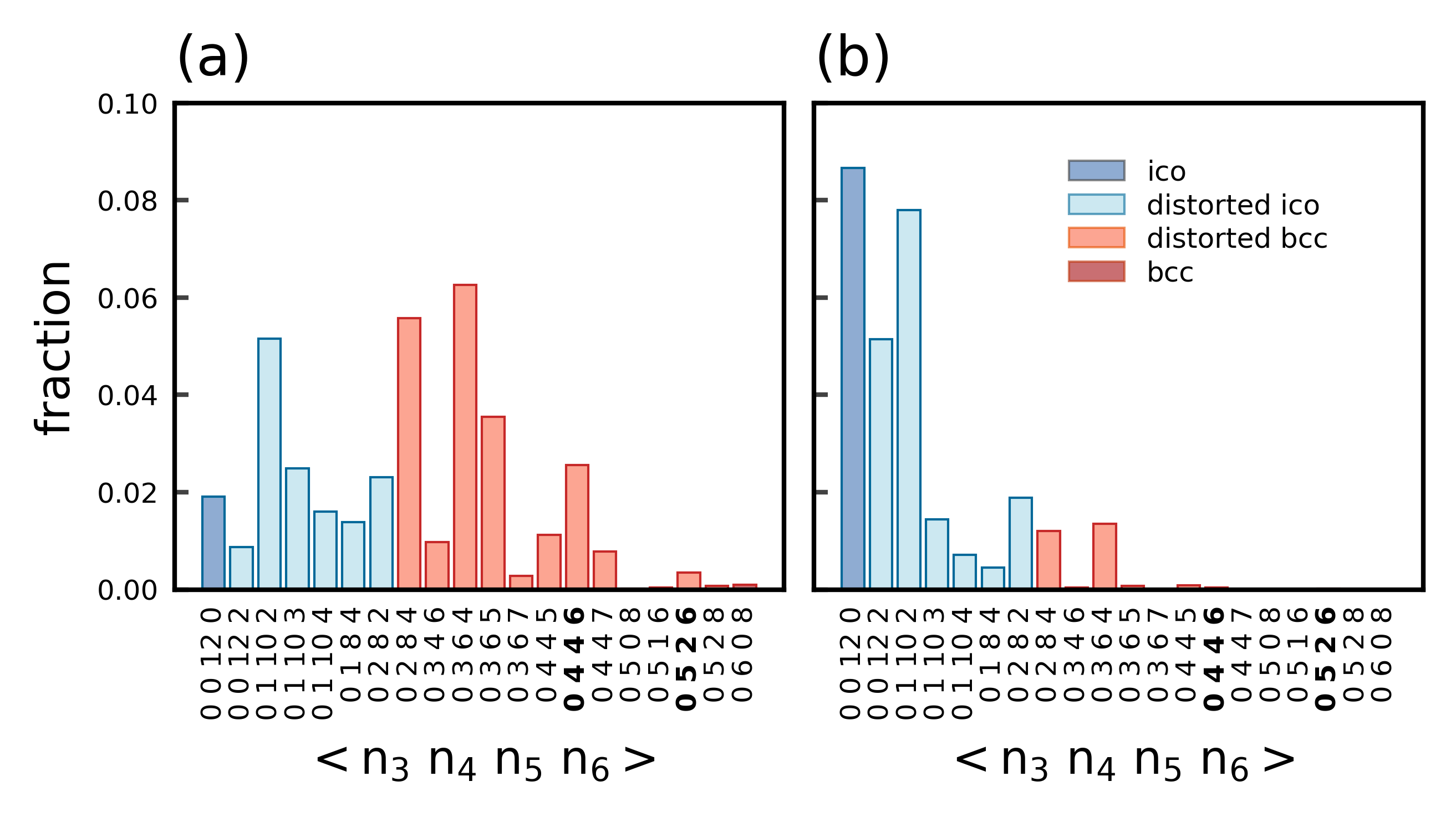}
    \caption{\label{fig:meam-vp}  Distributions of Voronoi polyhedra for liquid at 20\% undercooling using a (a) FS and (b) MEAM potentials. 
    Icosahedral polyhedra are shown in dark blue, distorted icosahedral polyhedra are in light blue and distorted bcc are in light red.
}
\end{figure*}

Due to the computational cost of TIS simulations it remains still unfeasible to combine them with energies and forces based on electronic structure calculations. On the other hand, empirical interatomic potentials are computationally efficient, but are often developed to reproduce specific reference properties, with a particular focus on crystalline bulk phases, and hence might suffer from limited transferability \cite{Lysogorskiy2019}. To evaluate the sensitivity of our results to the specific empirical potential used here, we have performed additional simulations with two further empirical potentials for Mo, a Finnis-Sinclair (FS)~\cite{Finnis1984, Ackland1987}  and a modified EAM (MEAM)~\cite{Lee2001} potential.

The simulations results obtained with the FS potential are very similar to the EAM ones.  The melting temperature is somewhat lower, $T_m = 2850$~K, and consequently the absolute values of the nucleation barriers and rates differ, but show the same trends and are in very good, qualitative agreement (see Supplementary Material~\cite{supplemental} for details).

The nucleation mechanism is the same in the two potentials, including the formation of a pre-ordered region in the liquid with the subsequent emergence of a bcc crystalline core.  Furthermore, the structural characteristics of the precursor are in excellent agreement, the distribution of VP obtained with the FS potential (shown in Fig.~S1 in the Supplementary Material~\cite{supplemental}) is nearly identical to the one shown in Fig.~\ref{fig:bulk_voronoi}.
Clearly, the results obtained with the FS potential are very much comparable and lead to the same assessment of the nucleation mechanism in Mo.

With the MEAM potential for Mo used in this work, nucleation and growth of bcc bulk from the undercooled liquid could not be observed.  Instead, an amorphous phase forms akin to a glassy state.
The failure of the employed MEAM potential to solidify into the crystalline bcc phase can be understood from the analysis of the undercooled liquid.  
In Fig.~\ref{fig:meam-vp} the distribution of VP are shown for the liquid at 20\% undercooling using the FS and MEAM potential.  While the distribution obtained with the FS potential is essentially the same as for the EAM potential in Fig.~\ref{fig:bulk_voronoi}, the distribution using the MEAM potential is clearly different.  The amount of icosahedra (dark blue) and distorted icosahedra (light blue) increases and the number of distorted bcc polyhedra (light red) decreases.  
Specifically, the absence of distorted bcc polyhedra impedes the formation of pre-structured liquid and pushes the system towards amorphization.
The MEAM potential does not accurately represent the heterogeneous nature of the undercooled liquid which is decisive in describing the nucleation process.~\cite{Russo2016}
A reason for this could be a  too strong contribution of the angular term in this MEAM potential.  

Comparing the three potentials, EAM, FS, and MEAM, it becomes apparent that a suitable empirical potentials needs to be able to correctly capture the structural features of both, the liquid and solid, phases of a material.  If this is fulfilled, the overall nucleation mechanism appears to be less sensitive to a particular empirical potential.

\section{Conclusions}

We analyze the nucleation mechanism during solidification in Mo which proceeds via two steps: the initial formation of  pre-ordered regions in the supercooled liquid and a subsequent nucleation of the crystalline bcc bulk phase within these regions.
The pre-ordered liquid region has a long lifetime and surrounds the bcc phase in the growing cluster even beyond the critical nucleus size.
The increase in bond-orientational order within this region creates a precursor that reduces the interfacial free energy by providing a diffuse interface between the liquid and crystalline core of the growing nucleus, thus facilitating the nucleation of the bulk phase.

While the density within the pre-ordered liquid remains constant for the growing clusters,
the bond-orientational order increases showing that this is the decisive factor that triggers the nucleation process.
Furthermore, the analysis of the structural composition of the pre-ordered region reveals clear differences from the liquid and solid phases, showing no icosahedral structures and a strong tendency towards (distorted) bcc like environments.  The structure of the pre-ordered region thus promotes the nucleation of a specific bulk structure and is key in the selection of the final polymorph. 

The structural heterogeneities in the undercooled liquid are very important in the crystallization process.  Indeed, we find that the nucleation mechanism strongly depends on  the ability of the interatomic potential to correctly capture  the structural hallmarks of the liquid. 
A complete absence of distorted bcc like environments and an abundance of icosahedral structures (as predicted by the employed MEAM potential) inhibit the formation of the precursor and, consequently, promote the formation of amorphous structures instead of crystalline bulk phases. This finding agrees well  with recent work showing that an increase in the structural difference between liquid and crystal enhances the glass-forming ability, and is linked with the suppression of crystal precursors.~\cite{Russo2016}

Comparing to other metallic systems, such as fcc Ni~\cite{DiazLeines2017}  and Al~\cite{Desgranges2007}, the overall nucleation mechanism is very similar and appears to be generally valid, in particular the formation of a precursor of high orientational order followed by the emergence of the  crystalline phase within the cluster core.  
The structure of the pre-ordered liquid is, however, inherently different for these systems, strongly indicating the key role of precursors in the selection of the final polymorph at the very early stages of nucleation.  This also implies that controlling the formation of specific bulk polymorph through seeding or environmental conditions is closely associated with controlling the structure and formation of the precursor zones in the supercooled liquid.

\section*{Supplementary Material}

See Supplementary Material for a details on the TIS setup, on nucleation barriers and rates, and on the comparison with the FS potential, which includes Refs.~\cite{Turnbull1950, Spaepen1994}.

\begin{acknowledgments}
S.M. acknowledges a scholarship from the International Max Planck Research School for Interface Controlled Materials for Energy Conversion.
We acknowledge financial support by the Deutsche Forschungsgemeinschaft (DFG) through project 262052203 and project 211503459 (C2 of the collaborative research center SFB/TR 103). The authors
acknowledge computing time by the Center for Interface-Dominated High Performance Materials
(ZGH, Ruhr-Universit\"{a}t Bochum).

\end{acknowledgments}

\vspace{10px}
\bibliographystyle{apsrev}

\bibliography{references}

\end{document}


\maketitle

\section{Transition interface sampling simulation details}

TIS simulations were carried out at three different undercoolings~($ \Delta T/T_m $), 20\%, 23\%, and 25\% for both, the embedded atom method~(EAM)~\cite{Zhou2004} potential and the Finnis-Sinclair potential~(FS)~\cite{Finnis1984, Ackland1987}.
The values of the order parameter $n_s$ that define the liquid and solid stable states and the TIS interfaces for both potentials, and for all the
undercoolings, are shown in table~\ref{tab:interfaces}. The interfaces were chosen in such a way that there is at least 10\% overlap in crossing probability between subsequent interfaces.

\begin{table}[htbp]
\caption{\label{tab:interfaces}
Interfaces and stable states for the EAM and FS potentials for undercoolings of 20\%, 23\%, and 25\%.}
\begin{tabular}{cccc}
\hline
Potential            & $\Delta T/T_m$ {[}\%{]} & Stable states {[}$n_s${]} & Interface positions {[}$n_s${]}                 \\ \hline
\multirow{7}{*}{EAM} & \multirow{3}{*}{20}     & \multirow{3}{*}{15, 400}  & 15,  25,  35,  50,  70,  80, 125, 140, 155,     \\  
                     &                         &                           & 170, 180, 190, 200, 230, 270, 300, 325,         \\  
                     &                         &                           & 350, 400                                        \\ \cline{2-4} 
                     & \multirow{2}{*}{23}     & \multirow{2}{*}{18, 200}  & 20,  40,  50,  55,  60,  75, 90, 100, 110, 120, \\  
                     &                         &                           & 125, 140, 150                                   \\ \cline{2-4} 
                     & \multirow{2}{*}{25}     & \multirow{2}{*}{20, 150}  & 20,  40,  50,  55,  60,  75, 90, 100, 110, 120, \\  
                     &                         &                           & 125, 140, 150                                   \\ \hline
\multirow{6}{*}{FS}  & \multirow{2}{*}{20}     & \multirow{2}{*}{15, 400}  & 15,  25,  35,  50,  70,  80, 125, 170, 230,     \\  
                     &                         &                           & 270, 300, 325, 350, 400                         \\ \cline{2-4} 
                     & \multirow{2}{*}{23}     & \multirow{2}{*}{18, 200}  & 18,  25,  35,  50,  70,  80, 110, 125, 132,     \\  
                     &                         &                           & 150, 170, 186, 200                              \\ \cline{2-4} 
                     & \multirow{2}{*}{25}     & \multirow{2}{*}{20, 150}  & 20,  40,  50,  60,  75,  90, 100, 110, 120,     \\  
                     &                         &                           & 125, 130, 140, 150                              \\ \hline
\end{tabular}
\end{table}

\section{Comparison with Finnis-Sinclair potential for Mo}

The calculated free energy barriers and rates for the EAM~\cite{Zhou2004} and the Finnis-Sinclair~\cite{Finnis1984, Ackland1987} potential extracted from the transition path ensemble are shown in table~\ref{tab:fscomparison}. The critical cluster sizes at each undercooling for both potentials are also included in the table.
\begin{table}[htbp]
\caption{\label{tab:fscomparison}
Rate constants, free energy barriers and critical cluster sizes at $25\%$, $23\%$ and $20\%$ undercoolings for EAM and FS potentials.}
 \begin{center}
\begin{tabular}{cccccc}
\hline
Potential            & $T_m$                    & $\Delta T/T_m$~(\%) & J ($10^{23}~(\mathrm{cm}^3)~(\mathrm{s}^{-1})$) & $\Delta G~$(eV)  & $n_s^*$  \\ \hline
\multirow{3}{*}{FS}  & \multirow{3}{*}{2850} & 20  & $6 \pm 2$                   & $3.82\pm0.02$           & 200 \\  
                     &                       & 23  & $4 \pm 0.8 \times10^{2}$       & $2.8 \pm 0.1$           & 148 \\ 
                     &                       & 25  & $8 \pm 6 \times10^{3}$       & $2.04 \pm 0.04$           & 115 \\ \hline
\multirow{3}{*}{EAM} & \multirow{3}{*}{3472} & 20  & $2.3\pm0.6$             & $4.41\pm0.02$ & 172 \\  
                     &                       & 23  & $1.4\pm0.3\times10^{2}$ & $3.7\pm0.2$   & 110 \\  
                     &                       & 25  & $4\pm1\times10^{4}$     & $2.1\pm0.1$   & 81  \\ \hline
\end{tabular}
 \end{center}
\end{table}
In classical nucleation theory (CNT), the free energy barrier of nucleation, $\Delta G$, is given by $\Delta G^* = \frac{16\pi\gamma^3}{3(\rho|\Delta\mu|)^2}$. $\gamma$ is the interfacial free energy for the formation of an interface between the solid and the liquid, $\Delta \mu$ is the difference in chemical potential between the two phases and $\rho$ is the density of the solid phase.
Using the approximations $|\Delta\mu| = \Delta H_f \Delta T/T_m$~\cite{Turnbull1950} where $\Delta H_f$ is the heat of fusion, and $\gamma = \gamma_m T/T_m$~\cite{Spaepen1994} with $\gamma_m$ being the interfacial free energy at melting temperature,
\begin{equation} \label{eq:dg}
	\Delta G^*( \Delta T) = \frac{BT^3}{\Delta T^2}
\end{equation}
where,
\begin{equation}
	B = \frac{16 \pi \gamma_m^3}{3 T_m (\rho \Delta H_f)^2}
\end{equation}
Hence by fitting the calculated free energy barriers to Eq.~\eqref{eq:dg}, the parameter $B$ can be estimated. The value of $B$ for the EAM and FS potentials is comparable.

\subsection{Critical cluster size}

Assuming a spherical nucleus, the number of atoms in the cluster is related to the radius by,
\begin{equation}
N = \rho_N \frac{4}{3} \pi r^3  \quad .
\end{equation}
$\rho_N$ is the number density. The critical cluster sizes in number of atoms for both potentials and the 
corresponding values of the radii are shown in table~\ref{table:radiifs}. Although $n_s^*$ is different for EAM and FS potentials for the undercoolings studied in this work, when comparing the critical cluster radii, the ratios are close to 1 and show only a maximum of 12\% difference at the undercooling of 25\%. This difference can stem from the difference between the two potentials in either $\gamma_m$ or $\Delta H_f$ and requires further investigation.

\begin{table}[ht]
\begin{center}
\caption{\label{table:radiifs}
Comparison of critical cluster size in EAM and FS.}

\begin{tabular}{ccccccccc}
\hline \\[-12px]
$\Delta T/T_m$ & \multicolumn{2}{c}{$\rho_\mathrm{N}$ [atoms/\AA$^3$]} & \multicolumn{2}{c}{$n_\mathrm{s}^*$} & $\frac{n_s^{\mathrm{FS}}}{n_s^{\mathrm{EAM}}}$ & \multicolumn{2}{c}{$r^*$ [\AA]} & $\frac{r^{\mathrm{FS}}}{r^{\mathrm{EAM}}}$ \\[5px] \hline 
                     & EAM        & FS         & EAM        & FS        &                        & EAM       & FS        &                        \\ \hline
20                   & 0.0606     & 0.0606     & 172        & 200       & 1.16                   & 8.78      & 9.24      & 1.05                   \\ 
23                   & 0.0608     & 0.0608     & 110        & 148       & 1.35                   & 7.56      & 8.34      & 1.1                    \\ 
25                   & 0.0609     & 0.0609     & 81        & 115        & 1.42                   & 6.82      & 7.67      & 1.12                   \\ \hline
\end{tabular}
\end{center}
\end{table}
 
\section{Voronoi polyhedra using FS potential}

 \begin{figure}[htbp]
  \centering
     \includegraphics[width=1.0\linewidth]{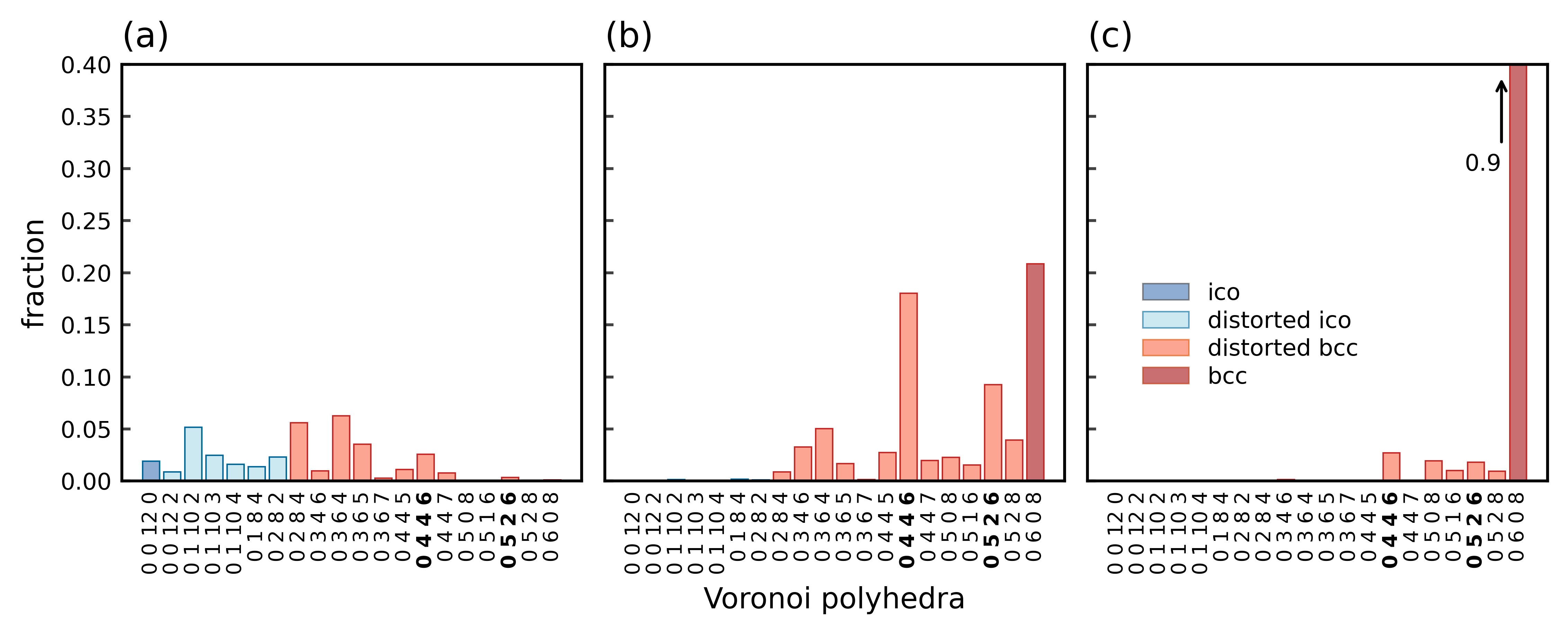}
     \caption{Comparison of polyhedra in (a) liquid, (b) pre-structured, and (c) solid particles for Finnis-Sinclair potential at 20\% undercooling.
     }
     \label{fig:poly_fs}
 \end{figure}
 
 A comparison of polyhedra in solid, liquid and pre-structured particles for Finnis-Sinclair potential at 20\% undercooling is shown in Fig.~\ref{fig:poly_fs}. The distributions for pre-structured particles were calculated using 100 configurations at the critical cluster size. 
 The distributions of VP obtained with the FS potential is essentially the same as the one obtained with the EAM potential.
 $\langle0~4~4~6\rangle$ and $\langle0~5~2~6\rangle$ are still the predominant pre-structured polyhedra. Similar to the EAM potential, polyhedra of the type $\langle0~6~0~8\rangle$, although distorted as compared to perfect bcc structure, is also present. The distribution of the Voronoi polyhedra remains similar at the range of undercoolings (20-25\%) studied.

\bibliographystyle{unsrt}
\bibliography{references}